\documentclass[12pt,onecolumn]{emulateapj}

 \makeatother

\def\vol#1  {{{#1}{\rm,}\ }}

\newcount\refno
\newcount\rfno
\def\eq{$^{\the\refno\ }$\advance\refno by 1}
\def\ad{\advance\rfno by 1}

\def\clock{\count0=\time \divide\count0 by 60
     \count1=\count0 \multiply\count1 by -60 \advance\count1 by \time
     \number\count0:\ifnum\count1<10{0\number\count1}\else\number\count1\fi}

\def\myputfigure#1#2#3#4#5%
{\hskip0.03\textwidth\vskip#5pt
\makebox[0pt]{\hskip#2in
\includegraphics[width=#3\textwidth]{#1}}\vskip#4pt\hfill}

\def\Gcm2{\rm G~cm^2}

\def\beq{\begin{equation}}
\def\eeq{\end{equation}}
\def\bea{\begin{eqnarray}}
\def\eea{\end{eqnarray}}

\def \date         {\ifcase\month \message{zero} \or
                    January \or February \or March \or April \or May \or June
                    \or July \or
                    August \or September \or October \or November \or
                    December \fi
                    \space\number\day, \number\year}

\begin{document}


\title{Hydromagnetics of advective accretion flows around black holes: Removal of angular momentum by large scale magnetic stresses}
\author{Banibrata Mukhopadhyay\thanks{bm@physics.iisc.ernet.in} and Koushik Chatterjee\thanks{kchatterjee009@gmail.com} \\ 
1. Department of Physics, Indian Institute of Science, Bangalore 560012\\
2. Department of Physics, Indian Institute of Technology Kharagpur, 
Kharagpur 721302}


\label{firstpage}

\begin{abstract}
We show that the removal of angular momentum is possible in the presence of
large scale magnetic stresses in geometrically thick, advective, sub-Keplerian 
accretion flows around black holes in steady-state, in the complete absence of $\alpha$-viscosity. The efficiency of 
such an angular momentum transfer could be equivalent to that of $\alpha$-viscosity 
with $\alpha=0.01-0.08$. Nevertheless, required
field is well below its equipartition value, leading to a magnetically stable disk flow.
This is essentially important in order to describe the hard spectral state of the sources,
when the flow is non/sub-Keplerian. We show in our simpler 1.5-dimensional, vertically 
averaged disk model that larger the vertical-gradient of azimuthal component of magnetic field,
stronger the rate of angular momentum transfer is, which in turn may lead to a
faster rate of outflowing matter. Finding efficient angular momentum transfer, in black hole
disks, via magnetic 
stresses alone is very interesting, when the generic origin of $\alpha$-viscosity is still
being explored.

\end{abstract}


\section{Introduction}

\cite{bp82} showed the possibility of energy and angular
momentum removal from a Keplerian accretion disk by the magnetic field 
lines that extend from the disk surface to large distance. In the framework
of infinite conductivity and self-similar model, they showed that the 
magnetic stresses can extract the angular momentum from a geometrically 
thin accretion disk, which helps matter to accrete, independent of the
presence of viscosity. Furthermore, they argued that such a mechanism 
is responsible for the observed jets/outflows from accreting sources,
when magnetic stresses convert a centrifugal outflow into a 
collimated jet. The disk matter has been argued to be outflowing 
through the outgoing field lines. The time evolution of axisymmetric, weak magnetic 
fields threading geometrically thin, Keplerian accretion disks with finite conductivity in 
a specific model framework was furthermore
investigated by \cite{lubow}, however without considering possible angular momentum transfer
by the magnetic field. 
On the other hand, in the presence of infinite 
conductivity, the magnetic field, in the same model framework which does not consider
the contributions from the magnetic stresses, would be kept on amplifying 
by the accretion of gas, till it 
stops the accretion (also see \citealt{spruit}). However, \cite{bish1} and \cite{bish2} showed that the radially
inward flow is possible
for plasma-$\beta>1$ and Prandtl number $\ge 1$, in the stationary
channel-type flows having small optical depth in the absence of turbulent viscosity, which also could exhibit electromagnetic outflows
for smaller Prandtl number. They showed that the large-scale field keeps drifting inward until 
a stationary state arises, when the magnetic, centrifugal, and gravitational
forces become comparable. This furthermore reveals the flow velocity profile differing significantly from the Keplerian profile.

The idea of exploring magnetic stress in order to explain astrophysical
systems was, in fact, implemented much earlier. For example, the solar wind 
was understood to have 
decreased Sun's angular momentum through the effect of magnetic stresses
(see, e,g., \citealt{waber-davis}), the proto-stellar gas 
clouds might have been contracted by magnetic effects (\citealt{mou-pal}). 
In the context of accretion disk, 
\cite{ozer-usov} and \cite{bland76} showed that the energy is possible to be extracted 
continuously by electromagnetic torques and twisted field lines. 
Furthermore, \cite{cao2} showed, by a linear stability analysis of the accretion disks,
that angular momentum is possible to be removed
by the magnetic torque exerted by a centrifugally driven wind. 
The same authors (\citealt{cao}) also discussed that
moderately weak fields can cause sufficient angular momentum loss, via a magnetic wind to 
balance outward diffusion in geometrically thin accretion disks.
However, plasma-$\beta$ has to be much smaller than unity to explain the tendency of strong 
flux bundles at the centers of disk to stay confined, as seen in numerical simulations.
Nevertheless, \cite{ogil} showed, by solving the local vertical structure of a thin 
accretion disk threaded by a poloidal magnetic field, the shortcoming of launching an outflow and 
suggested for an existence of additional source of energy for successful launching of the outflow.

However, observationally outflows/jets are mostly found to be emanating from 
the disk when it is in a hard state (e.g. \citealt{hard-out}), which is non/sub-Keplerian.
Note that jets appear
to be highly heterogeneous with velocities ranging from a few tens of 
million cm/sec to the escape velocity from the disk. Superluminal sources,
however, appear to exhibit jet velocity around the speed of light 
(\citealt{miley}).
The jets are found in disks around stellar mass black hole sources (e.g. 
GRS~1915+105) as well as supermassive black hole sources (e.g. M87).

Therefore, most of the modern models describing outflows/jets from the accretion disks
are based on sub-Keplerian, advective model when the flow has a significant 
radial velocity, unlike the Keplerian disk. For example, a class of self-similar, advection
dominated solutions was proposed by \cite{ny95}, in order to describe
bipolar outflows from black hole sources, e.g. Sgr~A*. Later on, such a class of advection 
dominated solution exhibiting outflows/jets was applied in many other contexts, e.g.
core-collapsing disks and gamma-ray bursts (\citealt{dimatteo}), 
low-radiative-efficiency nuclei of elliptical galaxies (\citealt{elli}).

In a different model framework, \cite{ct}, \cite{c99}, \cite{cm} described 
advective, sub-Keplerian accretion flows in order to explain outflows, quasi-periodic 
oscillations (QPOs) and spectral states in black hole sources. Furthermore, 
\cite{m03}, \cite{mg03}, \cite{rm10} described general advective accretion flows (GAAFs) 
around black holes and neutron stars and showed the effects of rotation of the black hole
on to the solutions. The last authors also included various cooling effects explicitly
and showed how the solutions get affected by the cooling properties. 

However, all the above models were formulated in the framework of Shakura-Sunyaev $\alpha$-viscosity 
(\citealt{ss73}), when the flows are assumed to be embedded with the plasma-$\beta>>1$.
Hence, the matter transport is assumed to be supported via turbulent viscosity, not by
large scale magnetic field, unlike that chosen by \cite{bp82}. Nevertheless, \cite{deb10} showed that
the transport is also possible in the presence of outflow in a 2.5-dimensional 
accreting system; it does not matter whether the outflow is magnetic or hydrodynamic. 
Note that outflows and even jets can also be formed in the absence of magnetic field.
This is likely to occur when the flow is radiation trapped and the accretion rate
is super-Eddington or super-critical (\citealt{love,begel06,febrika04,gm09}). 
 
The 2.5-dimensional accretion model, proposed by \cite{deb10}, will be complete
if the effects of (large scale) magnetic field is included therein. In that case, 
one presumably can explain outflow of matter plunging through the 
outgoing magnetic field lines more spontaneously, as \cite{bp82} did in the 
Keplerian framework.
To the best of our knowledge, so far there is no attempt to obtain a self-consistent
set of advective disk-outflow coupled accretion solutions in the presence of large scale 
magnetic field, which has lots of implication to explain low/hard state of sources.
This, however, has been discussed in some extent for circumstellar disks around young stars
(see, e.g., \citealt{konigl}), without discussing the detailed solutions of all the dynamical variables.
The present work steps forward in order to obtain such a set of solutions for black holes. 

In various numerical set ups, magnetohydrodynamic (MHD) simulations of accretion on to magnetized compact objects have 
already been explored. As examples, some of them considered axisymmetric systems in the presence of magnetosphere (\citealt{roma1}),
some other aimed at investigating advection of matter and magnetic field in the turbulent/diffusive disks (\citealt{roma2}),
when the field strength decreases due to reconnection and annihilation at a later time. 
Other groups, explored general relativistic magnetohydrodynamic (GRMHD) simulations of magnetically 
arrested accretion flows and outflows around black holes, for toroidally and poloidally dominated magnetic fields (\citealt{sasha1}).
They furthermore demonstrated the possible extraction of net energy from a spinning black hole via the Penrose-Blandford-Znajek mechanism
(\citealt{sasha2}). Moreover, there were radiation MHD/GRMHD simulations of accretion and outflows around black
holes, exploring three distinct flow phases including the radiatively inefficient phase which is similar to
the flows considered in the present work (\citealt{ohs,narayan}). Some of the MHD simulations investigated the reasons behind the 
variability in low angular momentum, underluminous accretion flows in the vicinity of a supermassive black hole (\citealt{proga}).
However, all these works, to the best of our
knowledge, considered the cases when any viscosity to be arisen from magnetorotational instability (MRI) leading 
effectively to the Shakura-Sunyaev $\alpha$ viscosity (\citealt{bh91}).

Here we plan
to investigate, semi-analytically, the effects of large scale magnetic field, with plasma-$\beta>1$ yet, 
on to the advective accretion flows in order to transport matter, however restricted to the simpler 1.5-dimensions.
Therefore, we consider the flow variables, averaged over the vertical coordinate,
to depend on the radial coordinate only. While the vertical equilibrium assumption 
corresponds to no vertical component of velocity, we choose the vertical
component of magnetic field to be non-zero. Although, in reality, a non-zero 
vertical magnetic field induces a vertical motion, in the platform of the present
assumption, any vertical motion will be featured as an outward motion. Nevertheless, whether
it is a vertical or outward transport, our aim is to furnish removal of angular 
momentum from the flow via magnetic stresses, leading to the infall of matter towards black holes.

The plan of the paper is the following. In the next section, we describe the 
set of magnetohydrodynamic/hydromagnetic equations, at the limit of very large Reynolds number,
as is the case in accretion disks, 
describing flow model. This is basically the set of Navier-Stokes equation,
but in the presence of magnetic shearing stresses (and Lorentz force), magnetic
induction equation, the condition for the absence of magnetic monopole, and 
finally the conservation of mass. Subsequently, we discuss the numerical solutions of
the set of equations in \S 3 and its implications. Finally, we summarize the
results along with a discussion in \S 4.

\section{Model hydromagnetic equations}

We describe optically thin, magnetized, viscous, axisymmetric, advective, vertically
averaged, steady-state accretion flow,
in the pseudo-Newtonian framework with the \cite{m02} potential.
The choice of the pseudo-Newtonian framework, for the present purpose, 
does not hinder any physics, compared to that would appear in the full general relativistic
framework. Hence, the equation of continuity, vertically averaged hydromagnetic
equations for energy-momentum balance in different directions
are given by
\begin{eqnarray}
\dot{M} = 4\pi x\rho h\vartheta,
\label{con}
\end{eqnarray}
\begin{eqnarray}
\vartheta\frac{d\vartheta}{dx}+\frac{1}{\rho}\frac{dP}{dx}-\frac{\lambda^2}{x^3}+F=\frac{1}{4\pi\rho}\left(B_x\frac{dB_x}{dx}+s_1\frac{B_zB_x}{h}-\frac{B_\phi^2}{x}\right),
\label{rad}
\end{eqnarray}
\begin{eqnarray}
\vartheta\frac{d\lambda}{dx}=\frac{1}{x\rho}\frac{d}{dx}\left(x^2W_{x\phi}\right)
+\frac{x}{4\pi\rho}\left(B_x\frac{dB_\phi}{dx}+
s_2 \frac{B_zB_\phi}{h}+\frac{B_xB_\phi}{x}\right),
\label{azi}
\end{eqnarray}
\begin{eqnarray}
{\rm where}~~~W_{x\phi}=\alpha(P+\rho\vartheta^2),
\nonumber
\end{eqnarray}
\begin{eqnarray}
\frac{P}{\rho h}=\frac{Fh}{x}-\frac{1}{4\pi\rho}\left(B_x\frac{dB_z}{dx}+s_3\frac{B_z^2}{h}\right),
\label{vert}
\end{eqnarray}
\begin{eqnarray}
\vartheta T\frac{ds}{dx}=\frac{\vartheta}{\Gamma_3-1}\left(\frac{dP}{dx}-\frac{\Gamma_1 P}{\rho}\frac{d\rho}{dx}\right)=Q^+-Q^-
=Q^+_{vis}+Q^+_{mag}-Q^-_{vis}-Q^-_{mag}
\label{tdss}
\end{eqnarray}
when we assume that the variables do not vary significantly in the vertical direction such 
that $\partial/\partial z\rightarrow s_i/z\sim s_i/h$, when $i=1,2,3$, which is indeed true in the disk flows.
Note that $s_1,s_2$ and $s_3$ are the degrees of scaling for
the radial, azimuthal and vertical components of the magnetic field respectively. 
As a consequence, the vertical component of velocity is zero.
Here $\dot{M}$ is the conserved mass accretion rate and the corresponding equation is 
the integrated version of the continuity equation, $\rho$ is the 
mass density of the flow, $\vartheta$ the radial velocity, $P$ the total pressure 
including the magnetic contribution, 
$F$ the force corresponding to the pseudo-Newtonian potential for rotating black holes, 
$\lambda$ the angular momentum per unit mass, $W_{x\phi}$ the viscous 
shearing stress written following
the Shakura-Sunyaev (\citealt{ss73}) prescription with appropriate modification (\citealt{mg03}), 
$h\sim z$, the half-thickness of the disk,
$s$ the entropy per unit volume, $T$ the (ion) temperature of the flow, $Q^+$ and $Q^-$
are the net rates of energy released and radiated out per unit volume in/from the flow
respectively (when $Q^+_{vis}, Q^+_{mag}, Q^-_{vis}, Q^-_{mag}$ are the respective
contributions from viscous and magnetic parts). We furthermore assume, for the present
purpose, the heat radiated out proportional to the released rate with the proportionality
constants $(1-f_{vis})$ and $(1-f_{m})$, respectively, for viscous and magnetic parts 
of the radiations. $\Gamma_1, \Gamma_3$
indicate the polytropic indices depending on the gas and radiation content in the flow (see, e.g., \citealt{rm10}
for exact expressions) and $B_x$, $B_\phi$ and $B_z$ are
the components of magnetic field.
Note that, the independent
variables $x$ and $z$ are the radial and vertical coordinates, respectively, 
of the flow, expressed in the units of $GM/c^2$, where $G$ is the Newton's
gravitation constant, $M$ the mass of the black hole and $c$ the speed of light.
Accordingly, all the above variables are made dimensionless, e.g. $\vartheta$ 
is expressed in the units of $c$. For any other
details, e.g. model for $Q^+_{vis}$, see the existing literature (\citealt{rm10,mukhraj10-2}), when
\begin{eqnarray}
Q^+_{vis}-Q^-_{vis}=\frac{\alpha f_{vis}(P+\vartheta^2\rho)\lambda}{x^2}.
\label{viss}
\end{eqnarray}

We furthermore do not consider the heat generated and absorbed due to the nuclear reactions
(\citealt{mc00,mc01}). This is to emphasize that all the variables appearing in the 
equations are assumed to be their respective vertically averaged quantities. 

Hydromagnetic flow equations must be supplemented by (for the present purpose, steady-state) 
equations of induction and no magnetic monopole, given by
\begin{eqnarray}
\nabla\times\vec{v}\times\vec{B}+\nu_m\nabla^2\vec{B}=0,
\label{indt}
\end{eqnarray}
\begin{eqnarray}
\frac{d}{dx}\left(xB_x\right)+s_3\frac{B_z}{h}=0,
\label{nomont}
\end{eqnarray}
when $\vec{v}$ and $\vec{B}$ are respectively the velocity and magnetic field vectors and 
$\nu_m$ is the magnetic diffusivity. On taking the ratio of the orders of the first to 
the second (diffusive) terms in equation (\ref{indt}), we obtain $L|\vec{v}|/\nu_m=R_m$,
when $L$ being the order of the length scale of the system. Hence, when the Reynolds number, $R_m$, is very large,
which is the case for accretion disks, the second 
term (which is associated with the magnetic diffusivity) in equation (\ref{indt}) can be 
neglected. However, this term can be rather important inside some localized regions in
certain astrophysical systems due to subtle reasons. Nevertheless, for the present purpose, 
for simplicity we 
will neglect this term throughout. Furthermore, as $\vartheta$ and $\lambda$ are assumed to be 
independent of the vertical coordinate, it is easy to check from the radial component of 
equation (\ref{indt}) that ${\partial B_z}/{\partial z}\rightarrow 0$ (and hence $B_z/h\rightarrow 0$). Therefore, 
the azimuthal and vertical components of equation (\ref{indt}), at large $R_m$, respectively lead to
\begin{eqnarray}
\frac{d}{dx}\left(\vartheta B_\phi-\frac{B_x\lambda}{x}\right)=0,
\label{ind1}
\end{eqnarray}
\begin{eqnarray}
\frac{d}{dx}\left(x\vartheta B_z\right)=0,
\label{ind2}
\end{eqnarray}
when the radial component of equation (\ref{indt}) turns out to be trivial.
Subsequently the equation (\ref{nomont}) reduces to
\begin{eqnarray}
\frac{d}{dx}\left(xB_x\right)=0.
\label{nomon}
\end{eqnarray}
Because of the choice of very large $R_m$ (ideal MHD), there is a perfect flux freezing
in the flow. Therefore, a steady advection of the vertical magnetic flux towards the center
may lead to the decrease of $\beta$, making it close to unity and further smaller, in a pure
axisymmetric flow, even if the initial $\beta$ was high. Hence, at some point,
the back reaction of the field will inhibit accretion, depending on the geometry of the 
field lines. Although the physics of this process is not captured by the equations above
and we also do not intend to discuss such physics, we will show below in \S 3 the effects
of higher magnetic fields, in particular the inner edge (around the critical radius), in the entire
flow structure. This is essentially to capture a situation after significant advection
done with a certain field geometry.

Henceforth, we will also neglect the second term in the parenthesis of equation (\ref{vert}).
The set of equations (\ref{con}), (\ref{rad}), (\ref{azi}), (\ref{vert}), (\ref{tdss}),
(\ref{ind1}), (\ref{ind2}) and (\ref{nomon}) is essentially the modified version of the set of advective accretion 
disk equations in the presence of a large-scale magnetic effect, which are otherwise 
discussed in the literature in absence of it. Generally, in order to understand the
hard state of accretion flows around black holes, the flow is assumed to be purely 
hydrodynamic with the consideration of a turbulent viscosity arisen due to a weak magnetic 
field, i.e. MRI (\citealt{bh91}). Although, MRI is a largely accepted idea,
so far, in order to explain the origin of turbulence in accretion disks, there are some subtle 
issues with it (e.g. \citealt{mahajan,mathew}) including its applicability in colder disks.
Therefore, transport of matter in disks 
is much more transparent through magnetic stresses, if the flows are embedded with
a large scale field. Such magnetic stresses are considered here on the right hand side
of the radial, azimuthal and vertical momentum balance equations. In addition, 
the magnetic heating due to abundant supply of magnetic energy and the annihilation of the 
magnetic fields (\citealt{kogan74,rai}), which effects may be small however, is considered in the energy equation
such that
\begin{eqnarray}
Q^+_{mag}-Q^-_{mag}=\frac{3f_m|\vec{B}|^2\vartheta}{16\pi x}.
\label{magg}
\end{eqnarray}
However, the other related terms in the momentum balance equations
are neglected, again in comparison with the remaining terms in the respective equations, for
the purpose of the present work.

Therefore, even in the 
absence of turbulent viscosity ($\alpha=0$) and hence viscous stresses, magnetic stresses
alone can help in transporting matter in the accretion flows. Such a consideration
of large scale magnetic field and hence transport via magnetic stress has,
although been considered in circumstellar disks around young stars (see \citealt{konigl}
for a recent review), not yet been considered for advective accretion flows around
black holes.

Question may arise, if the magnetic field with plasma-$\beta>1$ is adequate enough 
to describe infall of matter in order to explain observation. We will show in the 
next section that the large-scale magnetic field, even with a significantly 
large plasma-$\beta$, can describe advective accretion flows as efficiently as 
an $\alpha$ does. 

\subsection{Solution procedure}

We have seven equations (excluding the vertical momentum balance equation, which assures
the vertical magnetostatic balance) and seven variables: $\vartheta, \lambda, P,
\rho, B_x, B_\phi, B_z$, which we plan to solve along with the vertical magnetostatic balance condition. 
First, we plan to reduce $d\vartheta/dx$ in terms of other variables and the
independent variable $x$ alone (without any other derivatives), given by
\begin{eqnarray}
\frac{d\vartheta}{dx}=\frac{\frac{1}{F}\frac{dF}{dx}-\frac{3}{2x}-\frac{\rho}{2P}
\left(1+\frac{1}{\Gamma_1}\right)\left(\frac{\lambda^2}{x^3}-F+\frac{1}{4\pi\rho}
\left[\frac{s1 B_z B_x}{h}-\frac{B_\phi^2+B_x^2}{x}\right]\right)+\frac{\Gamma_3-1}
{2\vartheta\Gamma_1 P}\left[\frac{\alpha f_{vis}(P+\vartheta^2\rho)\lambda}{x^2}+
\frac{3f_m|\vec{B}|^2\vartheta}{16\pi x}\right]}{\frac{1}{\vartheta}
-\frac{\vartheta\rho}{2P}\left(1+\frac{1}{\Gamma_1}\right)}.
\label{dvdx}
\end{eqnarray}
As the advective accretion around black holes is necessarily transonic,
the flow must pass through a critical radius where
\begin{eqnarray} 
\vartheta=\vartheta_c=\sqrt{\frac{2\Gamma_1 P_c}{\rho_c(1+\Gamma_1)}},
\end{eqnarray}
when the variables with subscript `c' indicate the respective values at 
that critical radius and the numerator of equation (\ref{dvdx}) has to be zero
for a continuous solution. At the critical radius, we also prescribe 
\begin{eqnarray} 
B_{xc}=B_{yc}=B_{zc}=\sqrt{4\pi\rho_c}\frac{c_{sc}}{f_A\sqrt{3}},~~{\rm when}~~c_s={\rm sound~speed}\sim\sqrt{\frac{P}{\rho}},
\label{bcr}
\end{eqnarray}
so that the Alfv\'en velocity is a fraction of sound speed therein. Although this is a simpler 
prescription, other choices do not change the picture, being addressed in this work, qualitatively. 
Note that a steady MHD flow would normally have three critical points -- fast magnetosonic point,
Alfv\'en point and slow magnetosonic point -- of which the Alfv\'en point
is not a true critical point (\citealt{gammie}). The remaining two
physically important distinct critical points, corresponding to fast
and slow magnetosonic waves, collapse into a single point because of the assumptions made in equation (\ref{bcr}).
We typically choose
$f_A\sim 10-10^3$ in our various computations (see the figures). This is to capture a situation when magnetic pressure at the 
inner edge of the flow is not high enough to hinder radial infall of the matter, i.e. a situation with a weak back reaction 
of magnetic fields. Furthermore, from equation (\ref{nomon}),
we can write 
\begin{eqnarray} 
x_cB_{xc}={\rm constant}=C_0=xB_x
\label{nomonc}
\end{eqnarray}
which fixes the profile of $B_x$ throughout the flow.

These four conditions, along with the conditions
that $\lambda=\lambda_K$ (when $\lambda_K$ being the Keplerian angular momentum
per unit mass) and $\vartheta<< 1$ at the beginning of sub-Keplerian flow far away from the black hole,
i.e. outer boundary,
and $\vartheta\sim 1$ at $r\sim r_+$ serve as important
boundary conditions. 
Based on all the conditions, by solving the set of
seven coupled differential equations, we can obtain the profiles for all variables including those of
$B_y$ and $B_z$. Of course, then, one has to supply $M$, $\dot{M}$, $\alpha$, $\gamma$ and hence $f_{vis}$ and $f_m$ for
a flow. See \cite{rm10} for the solution procedure in details.

\section{Solutions}

Our main aim is to obtain the solutions of magnetized accretion flows. In other 
words, the aim is to understand how the large scale magnetic field (alone) can influence  
the mass transfer in accretion process, in particular in the the advective regime.
This is important in the light of our ignorance of the origin of viscosity
in accretion flows, which may be arisen from turbulence, as the molecular 
viscosity therein is well-known to be insignificant. Hence, our venture here is to 
investigate, if the large-scale magnetic field, however with large plasma-$\beta$,
can govern the same/similar transport of angular momentum as the well-known
$\alpha$-prescription does. Hence, we plan to understand the relative 
strengths between the magnetic stress tensors and the viscous stress tensors
in order to control advective accretion flows.

Our plan is to explore specifically two situations. (1) Flows with a relatively higher $\dot{M}$
and, hence, lower $\gamma$, modelled around stellar mass black holes: such flows may or may not form
Keplerian accretion disks. (2)
Flows with a lower $\dot{M}$ and, hence, higher $\gamma$, modelled around supermassive black holes:
such flows are necessarily hot gas dominated advective (or advection dominated) accretion flows.

\subsection{Accretion around stellar mass black holes}

We choose $M=10M_\odot$ and $\dot{M}=0.1\dot{M}_{Edd}$, when $M_\odot$ and 
$\dot{M}_{Edd}$ are the solar mass and the Eddington accretion rate 
respectively. However, this choice of $\dot{M}$
does not necessarily imply the flow to be purely Keplerian, rather advective, which is 
indeed, in general, under consideration. Such flows have 
temperature $T\gtrsim 10^9$K and $\rho\gtrsim 10^{-7}$gm/cc
(\citealt{monika,rm10}), which were extensively explored
in the context of the formation of shock in hot accretion disks and subsequent outflows and 
observed spectral states (\citealt{ct,c96}). A relation of $\gamma$ (and $\dot{M}$) with the 
ratio of the gas and radiation content of the flow and the corresponding variations have been
discussed by \cite{parikshit}. Therefore, following previous authors
and for the convenience of comparison of the previous results without magnetic fields, we choose 
$\gamma=1.335$ along with the intermediate $f_{vis}$ and $f_m$. 
Note that while the results depend on the sign of $s_2$, they do not
depend on $s_1$ and $s_3$.

\begin{figure*}
\centering
\includegraphics[scale=0.5]{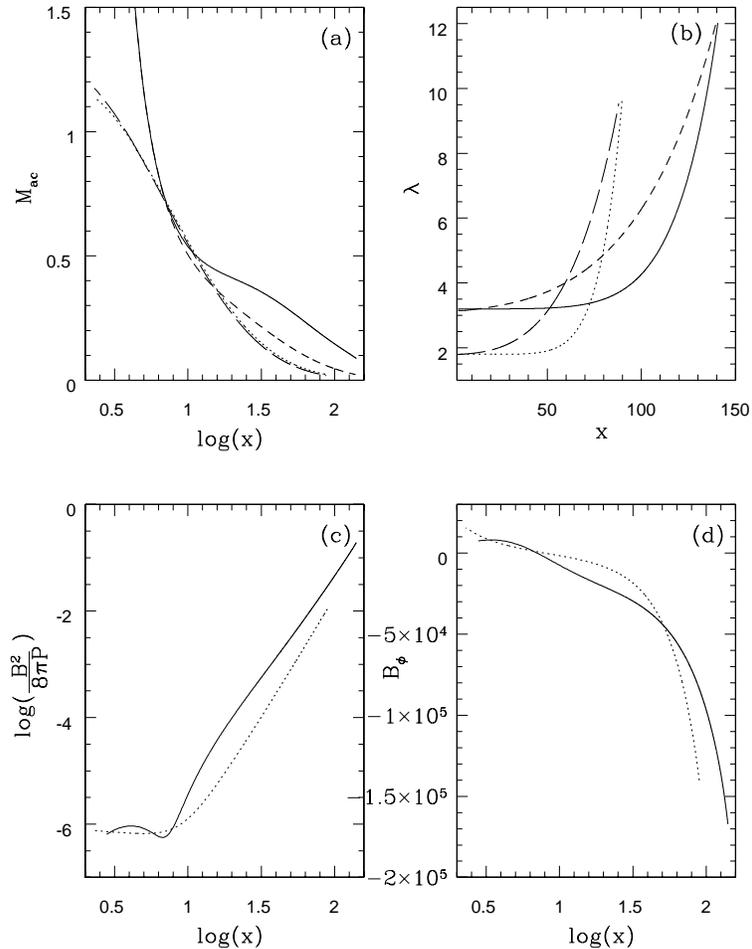}
\caption{(a) Mach number, (b) angular momentum per unit mass in $GM/c$, (c) inverse of 
plasma-$\beta$, 
(d) azimuthal component of magnetic field in Gauss, 
when solid and dotted lines are for magnetic flows around 
Schwarzschild ($a=0, \lambda_c=3.2$) and Kerr ($a=0.998, \lambda_c=1.8$) black holes respectively, 
and dashed and long-dashed lines are for viscous flows 
around Schwarzschild ($a=0, \alpha=0.017, \lambda_c=3.15$) and Kerr ($a=0.998, \alpha=0.012, \lambda_c=1.8$) 
black holes respectively. Other parameters are $M=10M_\odot$, $\dot{M}=0.1\dot{M}_{Edd}$,
$\gamma=1.335, f_{vis}=f_m=0.5, s_2=-0.5$.
}
\label{stmasm}
\end{figure*}

Figure \ref{stmasm} compares accretion flows (1) in the presence of 
large scale magnetic field, but in the absence of $\alpha$-viscosity:
magnetic flow, and (2) in the presence of viscosity, but in the absence of 
large scale magnetic field: viscous flow. 
Here we consider $B_\phi$ to have the same direction as of $\lambda$. 
It shows that the large
scale magnetic field $\sim 10^4-10^5$G, with its distribution in the inner edge of the flow defined by 
equation (\ref{bcr}), is adequately able to transport angular
momentum, as viscous flows do with $\alpha=0.017$ and $0.012$ respectively 
for nonrotating and rotating black holes. Figures \ref{stmasm}a,b show that the
disk sizes are the same for the respective viscous and magnetic sub-Keplerian flows
with the above mentioned respective $\alpha$-s. However, the transport of angular momentum
takes place faster in magnetic flows, where the flows become quasi-spherical at larger radii than 
at the radii they become in the respective viscous flows. Away from the black hole,
$B_\phi$ increases, which implies that the matter is prone to outflow 
from the outer region through the magnetic field lines extended outward
direction. In a self-consistent model, including the flow variation in the
vertical direction, the above features should have been appeared as the
increasing magnetic field with the vertical coordinate. Such a model
is planned to develop in future. In the present $1.5$-dimensional magnetic 
flow model,
when the magnetic stresses play the main role to remove angular momentum 
and hence to overcome the centrifugal barrier, as matter advances towards the 
black hole, the magnetic field decreases. Note that, as shown in 
Figs. \ref{stmasm}b,d, $\lambda$ and $|B_\phi|$
both decrease towards the black hole: to overcome large $\lambda$, the flow
needs a large $|B_\phi|$ and vice verse --- they are the self-consistent solutions
to each other. Note furthermore that the negative sign in $B_\phi$ in Fig. \ref{stmasm}(d) is 
due to our choice of decreasing $B_\phi$ with increasing $z$, i.e. negative $s_2$, in the 
outer edge of the flow (which results in the same trend of the flow almost throughput, except 
very inner region).
Figure \ref{stmasm}(c) shows that our chosen regime of magnetic flows, allowing 
a steady infall of matter, corresponds to a relatively high plasma-$\beta$ (actually inverse of $\beta$ in shown).
This furthermore renders lower magnetic pressures in respective flows compared to
their maximum allowed values based on the virial theorem/equipartition principle.
As discussed in the previous works (e.g. \citealt{m03}), a strong centrifugal
effect is depicted in the Mach number profiles in Fig. \ref{stmasm}a 
(featured as slowing down the matter at around $x=20-25$)
in the high angular momentum
flows around nonrotating black holes, compared to the low angular momentum
flows around rotating black holes.

\begin{figure*}
\centering
\includegraphics[scale=0.5]{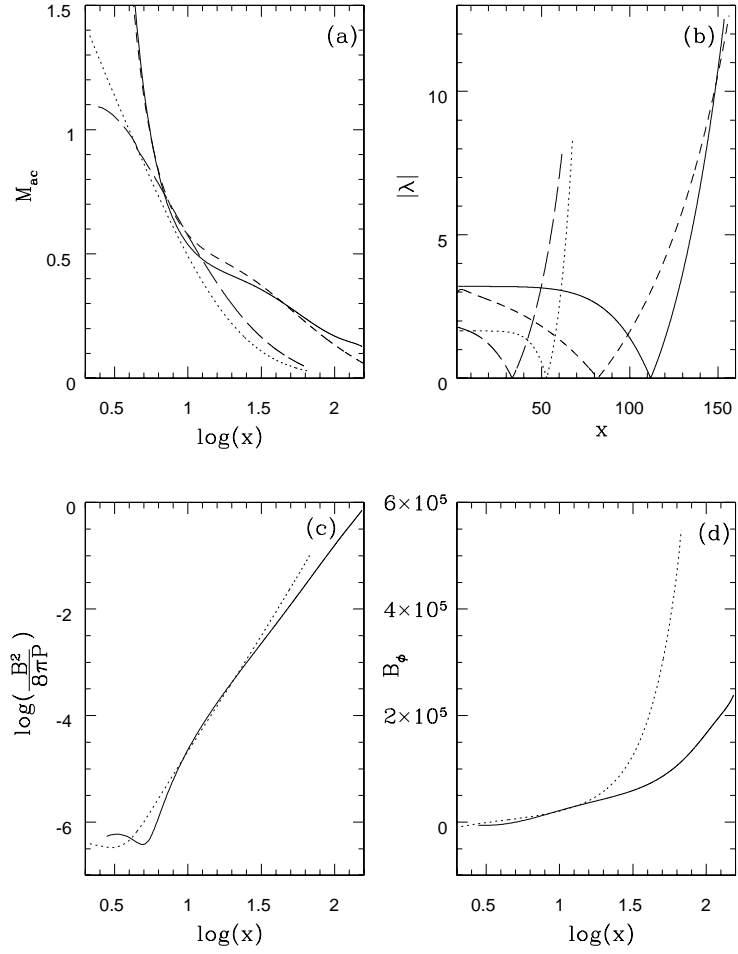}
\caption{
Same as in Fig. \ref{stmasm}, except $s_2=0.5$, when $\lambda_c=-3.2$ and $-1.65$
for nonrotating and rotating magnetic flows respectively and
$\alpha=0.08, \lambda_c=-3.08$ and $\alpha=0.056, \lambda_c=-1.74$ for nonrotating 
and rotating viscous flows respectively.
}
\label{stmasp}
\end{figure*}

Now we hypothesize that $B_\phi$ increases with the increasing $z$, i.e. positive $s_2$, at the 
outer edge of the flow (which results in the same trend of the flow almost throughout, except very inner region).
In this case, the right hand side of equation (\ref{azi}), for a magnetic flow, completely flips sign 
compared to the negative $s_2$ case.
Figure \ref{stmasp} shows that as the 
matter infalls, the toroidal component of magnetic field slows down the 
azimuthal motion of matter faster, making $\lambda=0$ and, subsequently, inverting
the orientation of $\lambda$. This is effectively due to the change in signs of magnetic
stress components: $B_xB_\phi$ and $B_zB_\phi$. Figure \ref{stmasp}b shows the variation 
of the magnitude of $\lambda$, as the matter falls in. $\lambda$ is positive 
far away from the black hole, but it is negative close to the black hole.
The location around $\lambda=0$ reveals
a trough-like region in the flow. Hence, in the either sides
of $\lambda=0$, there is a stronger centrifugal barrier which stores matter 
around $\lambda=0$ (due to the competition between radial and azimuthal flows). 
This region is prone to kick the matter out, producing outflows. Hence,
if $B_\phi$ increases with $z$ in the flow to start with, then
as matter advances towards the black hole, a ``potential well" forms to produce outflows.
Note, however, that very close to a rotating black hole, matter will be 
under the influence of the black hole completely and hence $\lambda$ cannot
have an opposite sign with respect to that of the black hole. Therefore, this 
solution is not valid very close to the rotating black hole. Indeed, the pseudo-Newtonian 
description is not applicable very close to the black hole.
Nevertheless, the above solution implies a possibility of having such an origin of outflows
in a magnetized accretion flow in the presence of a finite conductivity (when the field is not frozen 
with the matter, when the term associated 
with magnetic diffusivity in the induction equation is retained).
Note that the plasma-$\beta>1$ is maintained throughout the flows.

This furthermore motivates us to check with such a possibility in viscous flows
with viscosities $\alpha=0.08$ and $0.056$ respectively for 
nonrotating and rotating black holes. 
As shown in Figs. \ref{stmasp}a,b, the angular momentum profiles in the 
respective magnetic and viscous flows appear to be similar, which furthermore makes
the respective radial velocity profiles similar, unlike the previous cases, 
as shown in Fig. \ref{stmasm}, with the positive $\lambda$ throughout. In the previous
cases, the magnetic stresses are able to remove the angular momentum faster than 
the viscous stresses, in particular at a large distance 
from a Schwarzschild black hole, which is clearly
understood from Figs. \ref{stmasm}a,b. Although the same is true for a Kerr 
black hole, as the disk angular momentum itself is lower there, it does not
effectively create any impact on the Mach number profiles. 
However, due to the choice of larger 
$\alpha$, in the counterrotating cases, viscous stresses appear to be almost
equivalent to the magnetic stresses and hence the radial velocity profiles in either
of the respective flows appear similar. 

\begin{figure*}
\centering
\includegraphics[scale=0.5]{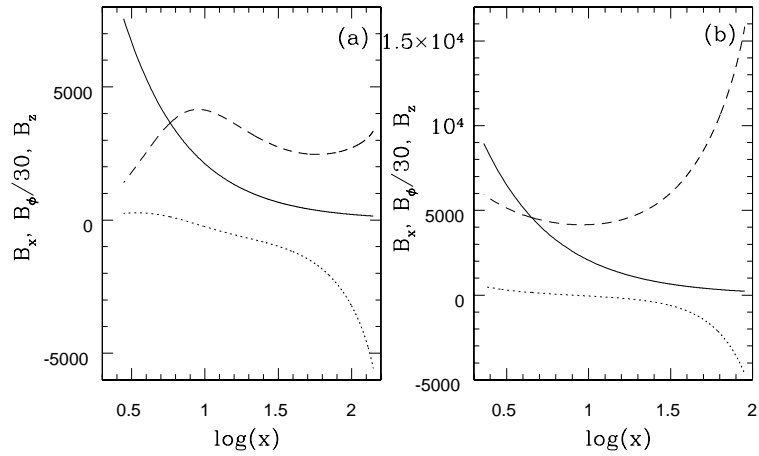}
\vskip-7cm
\caption{Components of magnetic field: $B_x$ (solid line),
$B_\phi$ but normalized by $30$ (dotted line), $B_z$ (dashed line), for (a) Schwarzschild 
magnetic flow of Fig. \ref{stmasm}, (b) Kerr magnetic flow of Fig. \ref{stmasm}.
}
\label{bcomp}
\end{figure*}

\begin{figure*}
\centering
\includegraphics[scale=0.5]{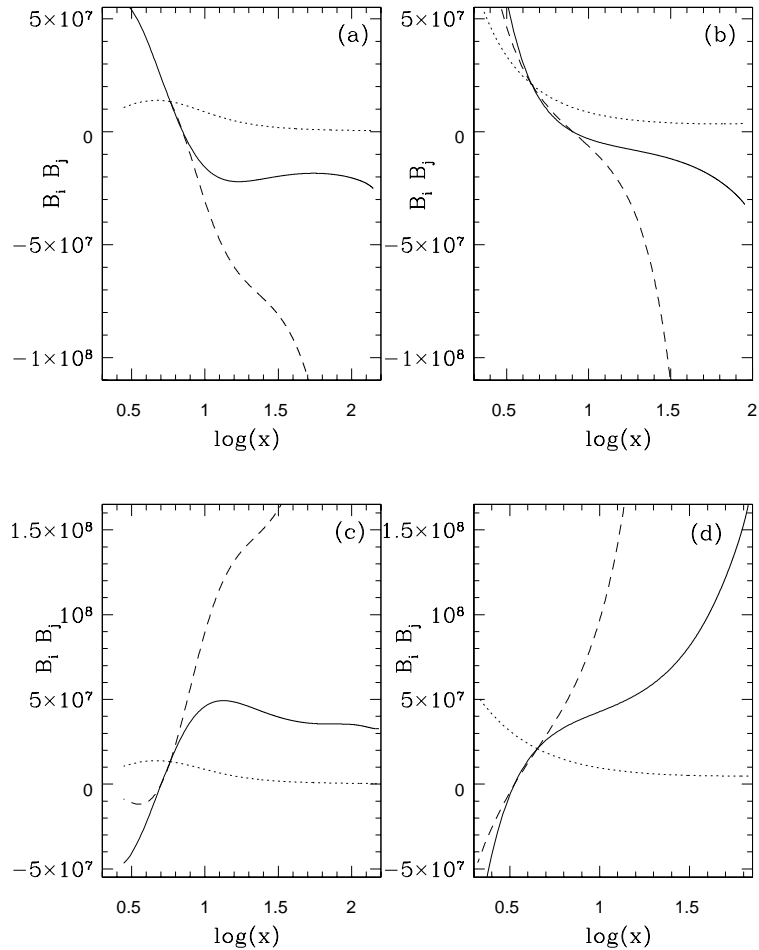}
\caption{Components of magnetic stress: $B_x B_\phi$ (solid line),
$B_x B_z$ (dotted line), $B_\phi B_z$ (dashed line), for (a) Schwarzschild 
magnetic flow of Fig. \ref{stmasm}, (b) Kerr magnetic flow of Fig. \ref{stmasm},
(c) Schwarzschild magnetic flow of Fig. \ref{stmasp},
(d) Kerr magnetic flow of Fig. \ref{stmasp}.
}
\label{stmasstr}
\end{figure*}

Let us now understand in more details, how the various components of magnetic stress are
responsible for inflow and/or angular momentum transfer therein. Figures \ref{bcomp}a,b
show the variations of various components of the magnetic field as functions of the radial coordinate,
around Schwarzschild and Kerr black holes, which are responsible for the various components of
magnetic stress tensor. The profiles of field components and their magnitudes are partly dependent on their
prescription given by equation (\ref{bcr}).
Figure \ref{stmasstr}a shows that the stress tensor component $B_x B_z$ around a Schwarzschild 
black hole increases almost throughout
as matter advances towards the black hole. This implies that the flow is prone 
to outflow through the field lines, which indirectly helps in 
removing the angular momentum, which furthermore renders its infall towards
the black hole. However, very close to the black hole, $B_x B_z$ decreases,
as indeed outflow is not possible in the near vicinity of the back hole, in
particular, once the matter passes through the (inner) sonic point. By this
radius, the flow angular momentum becomes very small which practically does
not affect the infall. The magnitude of $B_\phi B_z$ decreases till the 
inner region of accretion flow, implying the matter to be spiralling out 
and hence removing the angular momentum. A larger $|B_\phi B_z|$ at a larger radius implies a requirement of
the removal of larger $\lambda$ therein. This automatically emerges from the self-consistent
solutions of the set of equations. Nevertheless, close to the black hole,
this effect reverses, rendering infall. Finally, the magnitude of $B_x B_\phi$ 
increases at a large and a small distances from the black hole (except around
the transition radius), which helps infall, in the same way as the 
Shakura-Sunyaev viscous stress would do with the increase of matter pressure. 
However, at the intermediate zone,
the transfer of angular momentum through $B_x B_\phi$ reverses and a part of 
the matter outflows. At the Keplerian to sub-Keplerian transition zone,
flow/disk thickness increases, which effectively kicks the matter vertically,
showing a decrease of $B_x B_\phi$. Most of these features remain unchanged
for the flow around a rotating black hole, as shown in Fig. \ref{stmasstr}b. 
However, as a rotating black hole
renders a stronger/efficient outflow/jet, here, except at the inner zone,
$B_x B_\phi$ decreases throughout, which helps transferring the angular momentum
inwards and kicking the matter outwards. 
Nevertheless, such a flow does not exhibit a high $\lambda$ either so that does not 
necessarily require an increasing $B_x B_\phi$ to remove $\lambda$.
Figures \ref{stmasstr}c,d furthermore
confirm that the above properties are invariant for the cases of $B_\phi$
increased with increasing $z$. The only difference here is that $B_x B_\phi$
and $B_\phi B_z$ have opposite signs with respect to the previous cases, when the 
components of magnetic field considered here are their respective averaged values.
Note that all the components of magnetic stress tensor as functions of radius are 
determined by the associated components of magnetic field. The components of magnetic field
are, however, determined by solving the underlying set of equations self-consistently with their 
prescription at the inner edge of the flow for the regime of interest.

\begin{figure*}
\centering
\includegraphics[scale=0.5]{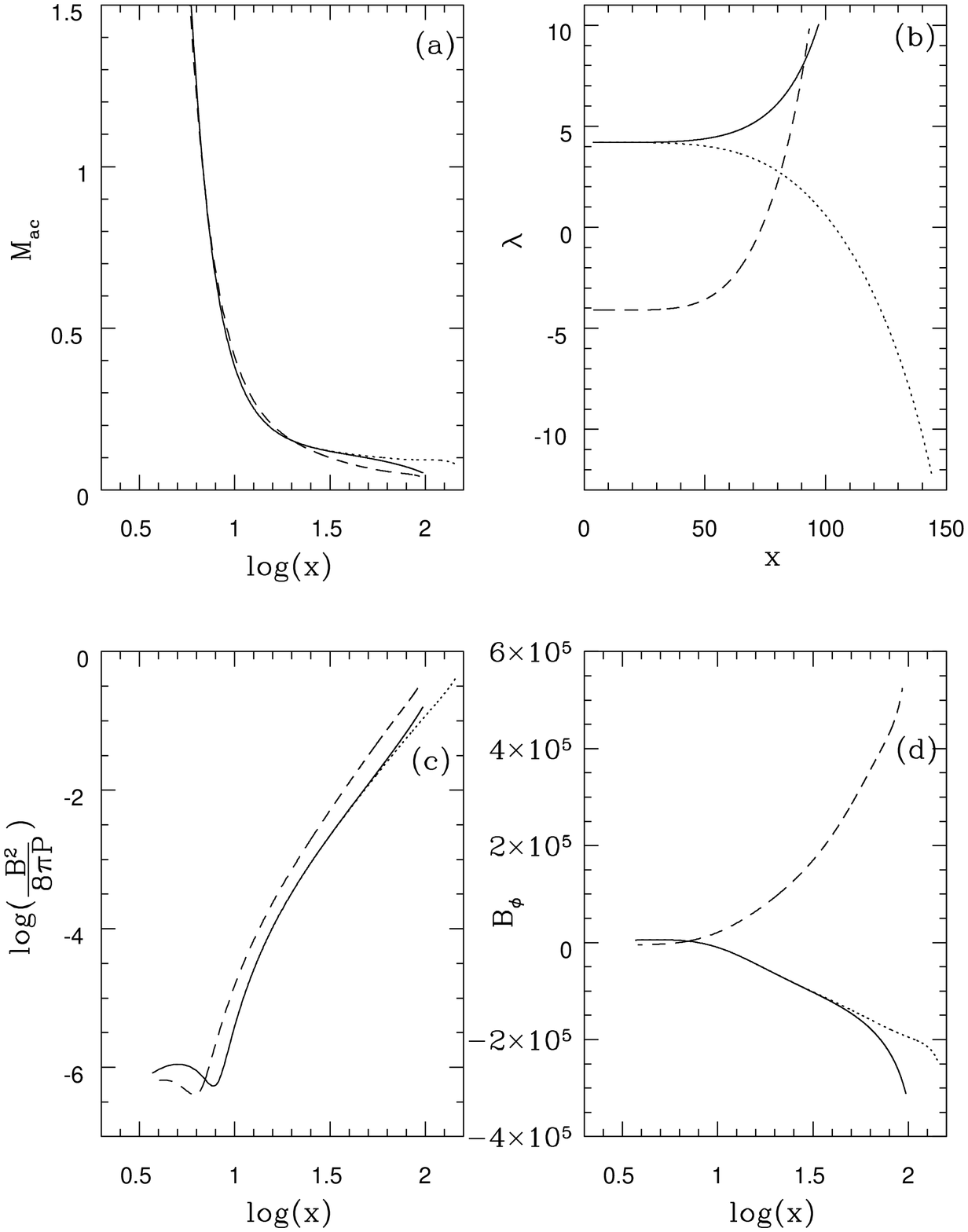}
\caption{
Same as in Fig. \ref{stmasm}, but all for magnetic flows around a
black hole with $a=-0.998$, when the solid line is for $s_2=-0.5, \lambda_c=4.2$
and the dotted and dashed lines are for $s_2=0.5$ with $\lambda_c=4.2$ and $-4.1$ respectively.
}
\label{stmasma}
\end{figure*}

Figure \ref{stmasma} compares three cases of magnetic flows. 
(1) A counter rotating disk throughout, when $B_\phi$ decreases with $z$ almost throughout (solid line).
(2) A disk having $B_\phi$ increasing with $z$ almost throughout, which is corotating 
far away, but counterrotating close to the black hole (dotted line).  
(3) A disk having $B_\phi$ increasing with $z$ almost throughout, which is
counterrotating far away, but corotating close to the black hole (dashed line).
Note importantly in the latter two cases that the increasing $B_\phi$ with disk height
induces the change of handedness of the disk during the infall of matter.
However, the profiles of
Mach number and $\beta$ practically appear similar in all the three cases.

\begin{figure*}
\centering
\includegraphics[scale=0.5]{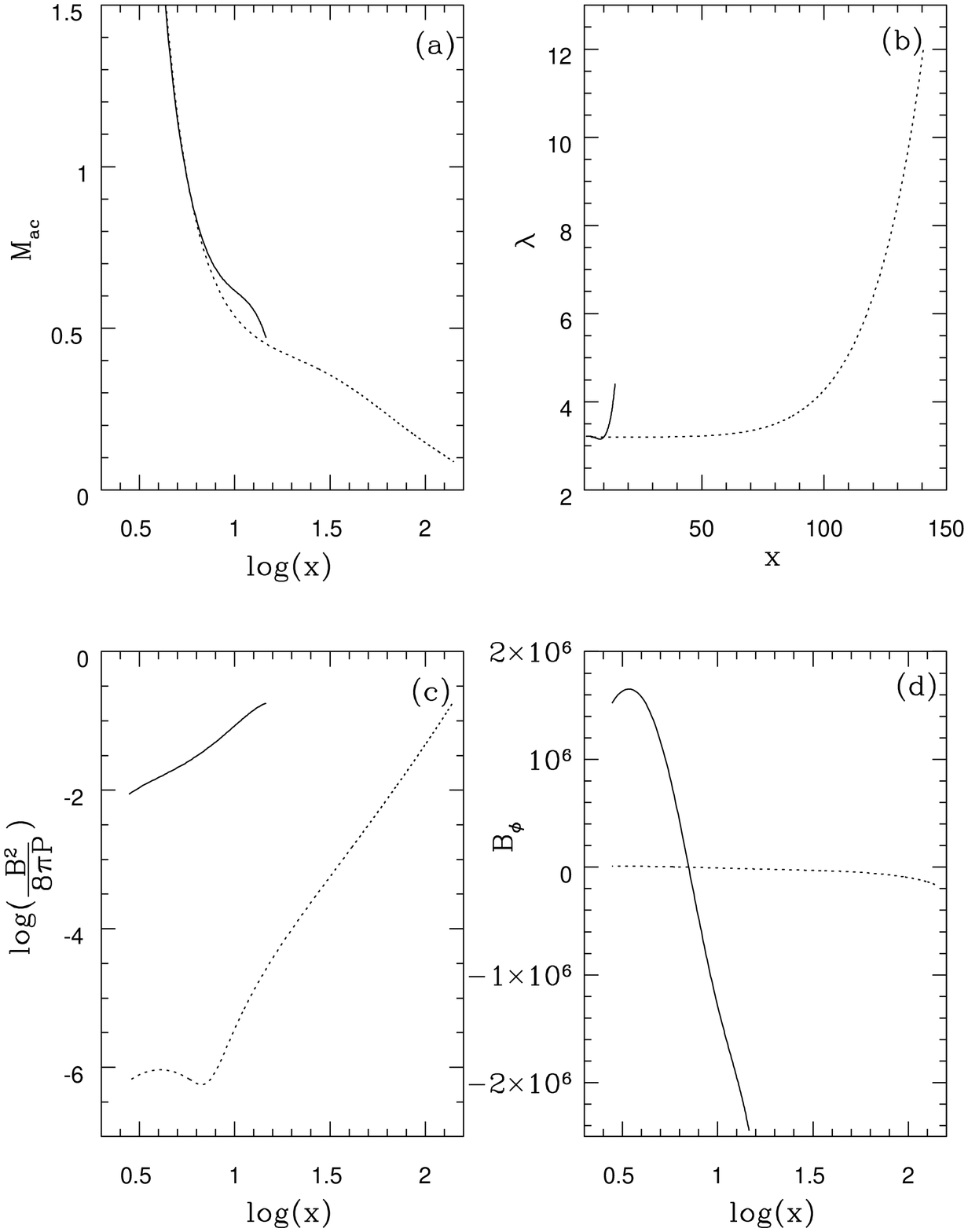}
\caption{
Same as in Fig. \ref{stmasm}, but comparing between the flows with
high (solid line) and low (dotted line) magnetic fields 
around a nonrotating black hole with $s_2=-0.5$.
}
\label{stmascom}
\end{figure*}

In Fig. \ref{stmascom}, we compare the disk hydrodynamics 
between the magnetic flows with high and low magnetic fields. 
As expected, a flow with the
higher magnetic field transports the angular momentum much faster,
leading to a smaller sub-Keplerian flow. In other words, 
in the presence of a higher magnetic field, when the magnetic stresses
are stronger, the Keplerian flow (when $\lambda=\lambda_K$) as well
as the boundary between the Keplerian and sub-Keplerian flows are able to
advance towards the black hole, shrinking the sub-Keplerian zone because of 
efficient angular momentum transfer.
Figure \ref{stmascom}c shows that at a given radius the magnetic 
pressure, and hence the Alfv\'en speed, is much larger in a flow with 
the higher magnetic field (but still $\beta>1$). Naturally, such a
high field magnetic flow would be equivalent to a viscous flow
with much larger $\alpha$, compared to the cases shown in Fig. \ref{stmasm}.  
An even higher magnetic field in the inner region would hinder any
infall due to backreactions.
Interestingly, in the radii close to the black hole, the sign of $B_\phi$ 
becomes distinctly opposite than the outer region in the high magnetic field case.
However, by this radius the required amount of angular momentum has already been 
transferred outwards in order to advance the matter close to the 
black hole and hence the change in sign of $B_\phi$ does not
create any physical impact onto the flow. If we vary the conditions chosen
in equation (\ref{bcr}), e.g., assume the components of magnetic field inequal,
the qualitative picture remains unchanged --- the magnetic stress in the presence of a large
scale magnetic field could adequately transfer angular momentum.
However, it is very important to note that if 
the strength of magnetic field and the corresponding value at the inner edge 
around the sonic radius would have been even higher, above a certain value, then the infall would no longer 
be possible. This is similar to the situation when above a certain value of 
$\lambda$ at the sonic radius in a given flow, the infall is no longer possible (see, \citealt{rm10}).

If the flow is gas pressure dominated with larger $\gamma$, then all the above basic features
remain the same. Be it radiation or gas dominated, large scale magnetic stresses,
yet $\beta>1$, can transport angular momentum as efficiently as the 
$\alpha$-prescription does. Nevertheless, below we discuss the effects of 
large scale magnetic field in the gas pressure dominated flows around a
supermassive black hole.

\subsection{Gas dominated accretion around supermassive black holes}

The supermassive black hole at the centre of our galaxy, Sgr~A*, 
presumably exhibits a gas pressure dominated, advection dominated, accretion flow.
This motivates us to undertake this case, when we choose
$M = 10^7M_\odot$ and $M = 10^{-4}M_{Edd}$, and hence an 
appropriate $\gamma=1.55$. We furthermore choose $f_{vis}=f_m=0.95$, when strongly advective 
matter hardly has a chance to radiate photons. However, such a flow may also arise 
around a stellar mass black hole, e.g. microquasars.

\begin{figure*}
\centering
\includegraphics[scale=0.5]{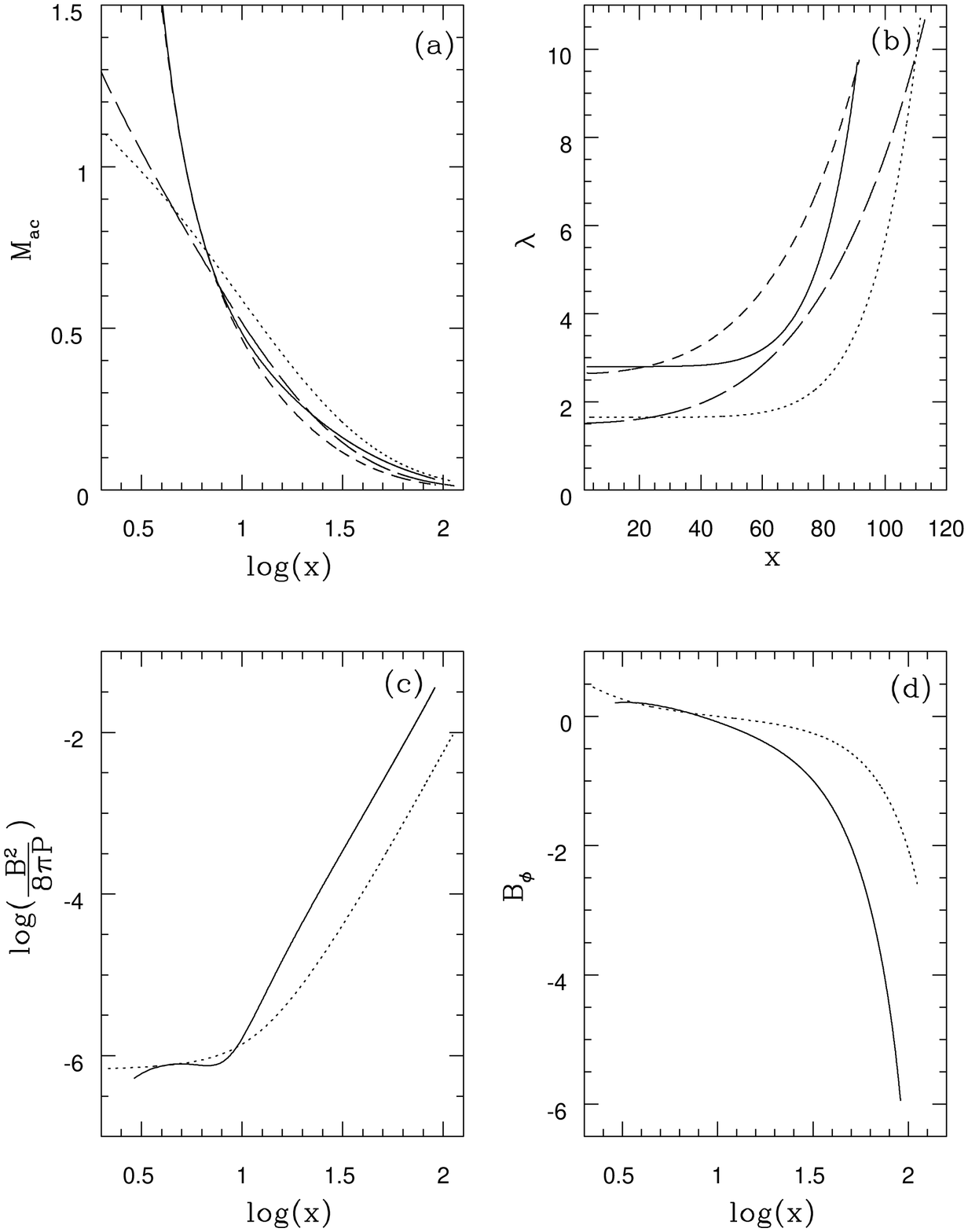}
\caption{(a) Mach number, (b) angular momentum per unit mass in $GM/c$, (c) inverse of 
plasma-$\beta$, 
(d) azimuthal component of magnetic field in Gauss, 
when solid and dotted lines are for magnetic flows around 
Schwarzschild ($a=0, \lambda_c=2.8$) and Kerr ($a=0.998, \lambda_c=1.65$) black holes respectively, 
and dashed and long-dashed lines are for viscous flows 
around Schwarzschild ($a=0,\alpha=0.011, \lambda_c=2.65$) and Kerr ($a=0.998, \alpha=0.0092, \lambda_c=1.52$) 
black holes respectively. Other parameters are $M=10^7M_\odot$, $\dot{M}=10^{-4}\dot{M}_{Edd}$, 
$\gamma=1.55, f_{vis}=f_m=0.95, s_2=-0.5$.
}
\label{sumasm}
\end{figure*}

The basic features in Fig. \ref{sumasm} are similar as those for 
the stellar mass cases, as shown in Fig. \ref{stmasm}. However, due to 
the gas dominance, and hence lower angular momenta, the Mach number profiles 
practically do not have any centrifugal barrier. Such flows are hotter, with $T\gtrsim 10^{11}$K, 
and more quasi-spherical, compared to the radiation dominated flows, when a very small part of
the dissipated heat can be radiated away. 
However, the most significant difference in these flows lies in their
low magnetic fields, compared to those discussed in \S3.1. This is due to
the largeness of black hole masses in these flows, which leads to a much larger
size of sub-Keplerian flows, when the dimensional flow size scales as $M$. As a result,
due to the law of equipartition, the magnetic field decreases significantly
compared to the cases of stellar mass black hole, as shown in Fig. \ref{sumasm}d
(as compared to those shown in Fig. \ref{stmasm}d). 

The magnetic flows around nonrotating and rotating, both the black holes, have their viscous 
counterparts with respective $\alpha=0.011$ and $0.0092$. This furthermore
confirms that even in gas dominated flows, the large scale magnetic field
is able to transfer the angular momentum as efficiently as the $\alpha$-prescription
does with the most plausible values of $\alpha$.

\begin{figure*}
\centering
\includegraphics[scale=0.5]{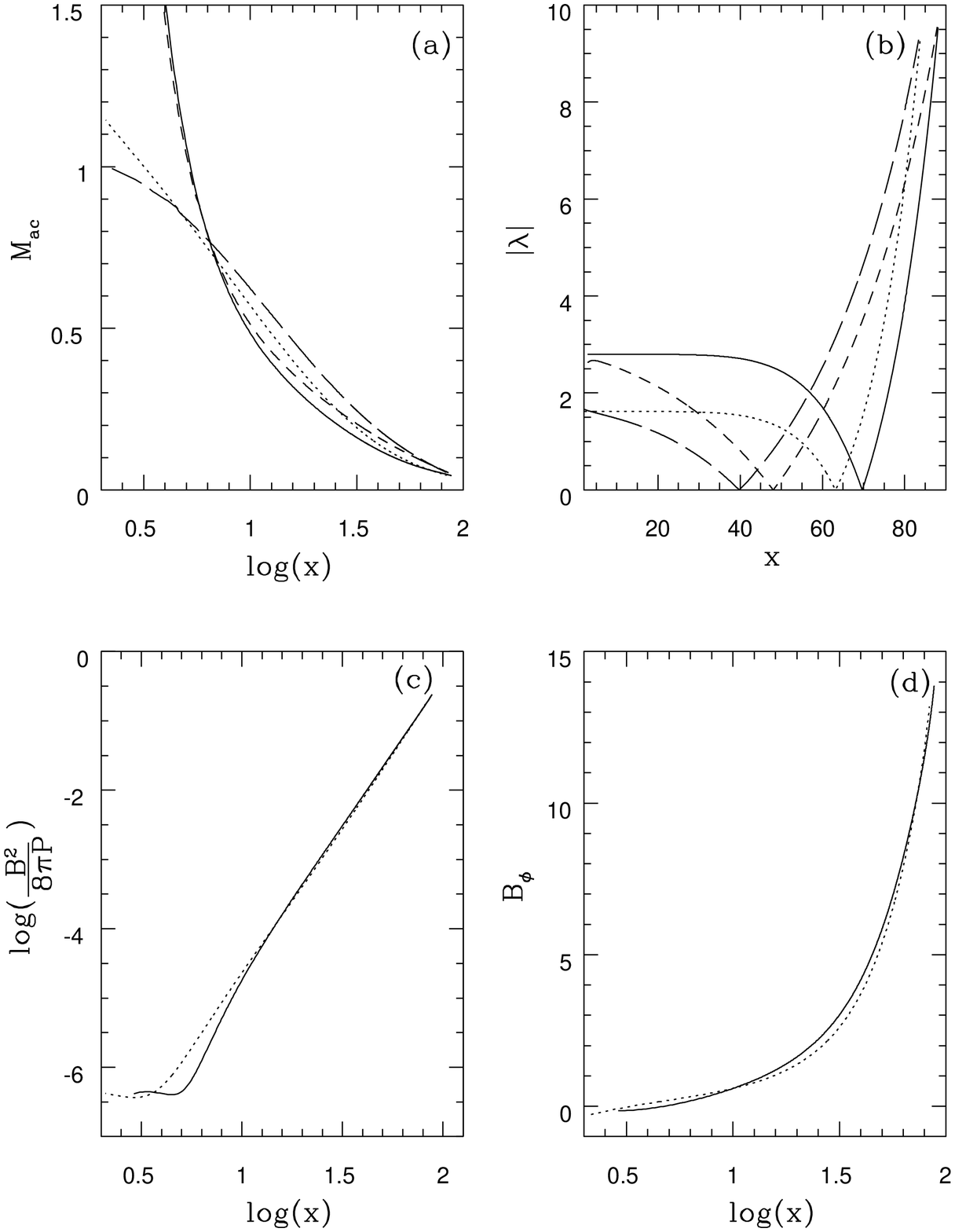}
\caption{
Same as in Fig. \ref{sumasm}, except $s_2=0.5$, when $\lambda_c=-2.8$ and $-1.62$ 
for nonrotating and rotating magnetic flows respectively and
$\alpha=0.075, \lambda_c=-2.65$ and $\alpha=0.07, \lambda_c=-1.6$ for nonrotating 
and rotating viscous flows respectively.
}
\label{sumasp}
\end{figure*}

Hypothesizing the increasing $B_\phi$ with increasing $z$ at the outer edge of 
the sub-Keplerian flow, we obtain the same results as those for 
stellar mass black hole accretion flows described above, except at much lower magnetic
fields. Like the stellar mass cases, here also a ``potential well" forms which is 
featured in Fig. \ref{sumasp}b, rendering the systems to have a zone for producing outflows.  
We do not repeat the detailed properties of it. Figures \ref{sumasp}a,b also show
the viscous flows resembling magnetic flows with
$\alpha=0.075$ and $0.07$ respectively for nonrotating and rotating black holes.

\subsection{Dependence on $s_1,s_2,s_3$}

As defined in \S2, $s_1,s_2,s_3$ parametrize the scaling of the variations of $B_x, B_\phi, B_z$ respectively in the 
vertical direction. This has be to considered because of our averaging the flows in the vertical
direction, while the variation of magnetic field in the vertical direction has not been neglected,
as has not been for $P$. 
Interestingly, the solutions practically do not depend on the choices of $s_1$ and $s_3$. However,
with the increase of the magnitude of $s_2$, which implies the increasing change of magnetic field with the 
vertical coordinate, the size of sub-Keplerian flow decreases. This is because, stronger the 
vertical variation of magnetic field, larger the change of the magnetic stresses in the vertical direction, 
on average faster the infall of matter is. This also argues for the faster rate of throwing the 
matter via outflows, when the outflows, in a more self-consistent 2.5-dimensional flow, are expected 
to plunge out via the magnetic field lines in the vertical direction. Hence, with the increase
of magnetic field in the vertical direction, the system becomes more prone to outflow matter. 
Subsequently, a faster rate of outflow renders a faster removal of angular momentum and 
hence a faster rate of infall. As a result, the flow could remain Keplerian with the aid of 
adequate mechanisms of angular momentum transfer, till further inner region of the flow. 
Hence, the boundary between the Keplerian and sub-Keplerian flows advances towards the black hole.
 
\subsection{Interconnection between advection and magnetic field}


As the current $\vec{J}$
in the conducting fluid with conductivity $\sigma$ and electric field $\vec{E}$ is known 
to be $\vec{J}=\sigma\left(\vec{E}+
\vec{v}\times\vec{B}\right)$, Faraday's law of induction in the steady-state for 
axisymmetric accretion disks considered here, as given by equation (\ref{indt}), can be
recalled as
\begin{eqnarray}
\nabla\times\frac{\vec{J}}{\sigma}-\nabla\times\left(\vec{v}\times\vec{B}\right)=0.
\label{ef}
\end{eqnarray}
For a flow with very large $R_m$ (when $\nu_m\propto \sigma^{-1}<<1$), the $z-$component of 
equation (\ref{ef}), averaged in $z$ and integrated over $\phi$, is given by
\begin{eqnarray}
\int\vartheta B_z xd\phi={\rm constant}=C,
\label{ef3}
\end{eqnarray}
when the constant can be identified as 
$C=d/dt\left(\int B_zds_{x\phi}\right)=d\Phi/dt$, where 
$ds_{x\phi}$ is the elementary surface area in the disk plane and 
$\Phi$ is the magnetic flux.
Therefore, equation (\ref{ef3}) fixes the relation between advection and 
$B_z$, and hence the magnetic flux in the accretion flow.
This also can be understood by recasting Faraday's law of induction into
\begin{eqnarray}
\nabla\times\left(\vec{E}+\frac{\partial \vec{A}}{\partial t}\right)=0,~~~{\rm when}~~~
\vec{B}=\nabla\times\vec{A},
\label{ind}
\end{eqnarray}
which furthermore argues for
\begin{eqnarray}
\vec{E}=-\nabla V-\frac{\partial \vec{A}}{\partial t}-\vec{C},
\label{e1}
\end{eqnarray}
when $V$ is the Coulomb potential and $\vec{C}$ is a constant vector. Hence, for a steady axisymmetric accretion flow
\begin{eqnarray}
E_\phi=\frac{J_\phi}{\sigma}+x\vartheta B_z=-C_\phi.
\label{e2}
\end{eqnarray}
Therefore for $R_m>>1$, $x\vartheta B_z$ is conserved. The constant $C_\phi$ or $C$ can be
fixed from a given boundary condition. In really, however, the flow is not expected to be 
purely axisymmetric and, hence, $C_\phi$ or $C$ can also be mimicked as the contribution 
from non-axisymmetry.
Earlier \cite{lubow} discussed a model of geometrically thin
accretion flows in the presence of weak magnetic field, 
but assuming $C_\phi=0$ which is not true in general.

The constraint on advection arisen in equation (\ref{e2}) is clearly visible in the 
velocity profiles in Fig. \ref{stmasm}a with respect to the variations of the vertical component of 
magnetic field shown in Fig. \ref{bcomp}. For the magnetic flow around a Schwarzschild black hole,
$\vartheta$ first increases steadily with the decrease of $B_z$ at larger radii and 
subsequently matter tends to slow down due to centrifugal effect 
(with the relative decrease of infall rate 
$d\vartheta/dx$) with the increase of $B_z$ until $x=10$. Finally, matter
plunges into the black hole steadily with a sharp decrease of $B_z$. For a rotating 
black hole, however, $\vartheta$ steadily increases with the steady decrease of $B_z$
almost throughout. Nevertheless, very close to the black hole, $B_z$ slightly increases
due to the spin effect of black hole, decreasing $d\vartheta/dx$ slightly.
This hints the power of black hole's spin to plunge the matter out.

\section{Discussion and Summary}

We have discussed the power of large scale magnetic field in advective accretion flows 
around black holes in oder to transport angular momentum, enabling infall of matter. 
In a simpler Keplerian, self-similar model framework, such an investigation
had been initiated  by \cite{bp82} long ago, and in the cases of circumstellar disks around 
young stars (e.g. \citealt{konigl}), such an approach has been explored. However, it remained
unexplored, to the best of our knowledge, in the advective accretion disk around 
black holes, when it may exhibit a hard spectral state, until this work. Note that, often, only hard spectral states
of disks are associated with the outflows/jets.

We have found that the flows
with plasma-$\beta>1$ exhibit adequate magnetic transport --- as efficient as
the $\alpha$-viscosity with $\alpha=0.08$ would do. This is interesting as the origin 
of $\alpha$ (and the corresponding instability and turbulence) is
itself not well understood. The maximum required large scale magnetic field is a few factor
times $10^5$G in a disk around $10M_\odot$ black holes and $\sim 10$G in a 
disk around $10^7M_\odot$ supermassive black holes. The presence of such a field, 
in particular for a stellar mass black hole disk when the binary companion supplying 
mass is a Sun-like star with the magnetic field on average $1$G, may be 
understood, if the field is approximately frozen with the disk fluids (or the supplied fluids from the 
companion star remain approximately frozen with the magnetic field) or disk fluids exhibit
large Reynolds number. Indeed, all the present 
computations are done at the limit of large Reynolds number, as really is the case in accretion flows,
such that the term associated with the magnetic diffusivity in the induction equation can be 
neglected. The size of a
disk around supermassive black holes is proportionately larger compared to
that around a stellar mass black hole. Hence, from the equipartition theory, indeed the 
magnetic field is expected to be decreased here compared to that around stellar mass
black holes.

Is there any observational support for the existence of such a magnetic field, as
required for the magnetic accretion flows discussed here?
Interestingly, the polarization measurements in the hard state of Cyg~X-1 imply
that it should have 
at least $10$mG field at the source of emission (\citealt{science}). In order to
explain such high polarization, a jet model was suggested by \cite{zd}, which 
requires a magnetic field $\sim (5 - 10)\times 10^5$G at the base of jet and hence 
in the underlying accretion disk. Also, the magnetic field in the inner region of accretion
disks around more than a dozen black holes has been found to be very high,
based on a model relating the observed kinetic power of relativistic jet
to the magnetic field of acretion disks (\citealt{arx}).

Different components of magnetic stress tensor have different roles: $B_x B_\phi$ 
controls the infall in the disk plane, whereas $B_\phi B_z$ renders the flow to spiral outwards and, hence, outflow.
Moreover, $B_xB_z$ helps to kick the matter out vertically.
Larger the field strength, larger is the power of magnetic
stresses. Interestingly, the magnitude of magnetic field decreases, as the steady-state
matter advances towards the black hole. This is primarily because $B_\phi B_z$ 
(and also $B_x B_\phi$ for a rotating black hole) decreases inwards almost entirely in order to induce outflow. 
This furthermore reveals a decreasing $|B_\phi|$ as the output of self-consistent
solutions of the coupled set of equations, which is reflected in the $|\vec{B}|$ profile.

In the present computations, we have assumed the flow to be vertically averaged
without allowing any vertical component of the flow velocity (but keeping all the 
components of magnetic field). The most self-consistent approach, in order
to understand vertical transport of matter through the magnetic effects which in turn leads
to the radial infall of rest of the matter, is considering the flow to be moving in the vertical direction
from the disk plane as well. Such an attempt, in the absence of magnetic effects,
was made earlier by \cite{deb10} in the model framework of coupled disk-outflow systems.
In such a framework, the authors furthermore showed that the outflow power of the
correlated disk-outflow systems increases with the increasing spin of black holes.
Our future goal is now to combine that model with the model of present work, so
that the coupled disk-outflow systems can be investigated more self-consistently and 
rigorously, when the magnetic field plays indispensable role in order to generate
vertical flux in the three-dimensional flows.

\section*{Acknowledgments}

B.M. acknowledges partial support through research Grant No. ISRO/RES/2/367/10-11
and K.C. thanks the Academy Summer Research Programme for offering him a
Fellowship to pursue his internship in IISc when most of the 
calculations of this project were done. 
The authors thank Omer Blaes of UCSB, Prateek Sharma of IISc and Alexander Tchekhovskoy of 
UCB and LBNL for many illuminating discussions and the anonymous referee for many useful suggestions.
The authors are also thankful to A. R. Rao of TIFR
for providing useful references which reveal observational inference of large scale 
magnetic field in an accretion disk.

\label{lastpage}

\end{document}